\documentstyle[12pt]{article}
\input epsf

\begin{document}\setlength{\unitlength}{1mm}

\def\question#1{{{\marginpar{\small \sc #1}}}}
\newcommand{\QCD}{{ \rm QCD}^{\prime}}
\newcommand{\MSSM}{{ \rm MSSM}^{\prime}}
\newcommand{\eq}{\begin{equation}}
\newcommand{\en}{\end{equation}}
\newcommand{\bino}{\tilde{b}}
\newcommand{\tsquark}{\tilde{t}}
\newcommand{\gluino}{\tilde{g}}
\newcommand{\photino}{\tilde{\gamma}}
\newcommand{\wino}{\tilde{w}}
\newcommand{\mtilde}{\tilde{m}}
\newcommand{\higgsino}{\tilde{h}}
\newcommand{\gsi}{\,\raisebox{-0.13cm}{$\stackrel{\textstyle>}
{\textstyle\sim}$}\,}
\newcommand{\lsi}{\,\raisebox{-0.13cm}{$\stackrel{\textstyle<}
{\textstyle\sim}$}\,}
\rightline{CERN-TH/98-269}
\rightline{hep-ph/9808439}
\rightline{August 1998}
\baselineskip=18pt
\begin{center}
{\large The Electroweak Phase Transition in the MSSM}\\
\vspace*{0.2in}
{ Marta Losada\footnote{On leave of absence
from  Universidad Antonio Nari\~{n}o, Santa Fe de
Bogot\'a, Colombia.}\\
{\it CERN Theory Division \\ CH-1211 Geneva 23, Switzerland}}\\

\vspace{.1in}
\end{center}
\vskip  0.1in  

{\small
 {\abstract The construction of an effective 3D theory at high temperatures
for the MSSM as a model of electroweak baryogenesis is discussed. The analysis for a single light scalar field shows, that given the experimental constraints,
there is no value of the Higgs mass for which a sufficiently strong first-order
phase transition is obtained.  A precise determination of the 3D parameters
of the effective theory for the case of a light right-handed stop allows us to obtain an upper bound on
the masses of the lightest Higgs and right handed stop using the two-loop 
effective
potential. A two-stage phase transition persists for a small 
range of values of $m_{\tilde{t}_{R}}$.}

\section{Introduction}
The study of the electroweak phase transition is motivated in part by
the idea of generating the baryon asymmetry of the Universe (BAU) at the
electroweak scale. As Sakharov \cite{sakharov} pointed out, the necessary ingredients for baryogenesis are: CP violation, non-equilibrium, and
baryon-number violation. The Standard Model has all of the
necessary features for the production of the BAU. In the Standard Model
 baryon number
is violated non-perturbatively through sphaleron processes.
At zero temperature the tunnelling rate of the anomalous processes is strongly
suppressed. However,  at high temperatures the situation changes drastically;
comparing the rate of sphaleron processes to the expansion rate
of the Universe, Kuzmin et al. \cite{kuzmin} showed that these processes are in equilibrium for
temperatures above the electroweak phase transition.
Thus, unless a net $B$-$L$ asymmetry is produced above the electroweak scale,
sphaleron processes will wash out the $B$ asymmetry. 
We can consider alternatively that the production of the baryon asymmetry takes place at the electroweak scale precisely because of the
 anomalous electroweak processes.  The generation of the baryon asymmetry is a non-trivial problem, which involves non-equilibrium dynamics. Here we focus instead 
on avoiding the elimination of the produced baryon asymmetry immediately
after the electroweak phase transition.
 At the electroweak scale, typical weak reaction rates are faster that the expansion rate of the 
Universe. This implies that  another mechanism for
a departure from thermal equilibrium is needed. The mechanism that has
been commonly employed relies on a sufficiently strong first-order phase transition
at the electroweak scale.
In addition, as weak-scale physics is involved, experimental constraints
can indicate whether, within a specific model, the requirement of preserving the asymmetry is satisfied.
 The rate of sphaleron transitions in the broken phase is proportional to
\begin{equation}
\Gamma \propto T e^{E_{sph}/T},  
\label{sphrate}
\end{equation}
where  $ E_{sph} = {m_{W}\over \alpha_{W}},$  is the sphaleron energy and $m_{W} = {1\over 2} g\phi$ is the gauge-boson mass. For the BAU to survive, baryon-number violation must be turned off
after the phase transition. This implies that the exponent in eq.
(\ref{sphrate}) must be large just after the phase transition\footnote{In our analysis we will be working in the case of a single light-scalar doublet at
the phase transition and so it is appropriate to use the constraint from
eq. (\ref{EsphT}).}
\begin{equation}
{E_{sph}\over T} > 45 \hspace{.3in}\rightarrow {v(T_{c})\over T_{c}} \gsi 1,
\label{EsphT}
\end{equation}
where $v(T_{c})$ is the expectation value of the Higgs field at the
critical temperature.

The effective potential is the quantity used in the analysis of the
phase transition to determine the order of the transition, critical
temperature, latent heat,
etc. For a precise determination of the effective potential the contributions
from all of the particles that receive mass terms from the scalar Higgs 
field should be included.
A generic expression for the effective potential, using the high-temperature
expansion, assuming a single light-scalar field at the phase transition is given by

\begin{equation}
V(\phi,T) = D(T^{2} - T_{o}^{2})\phi^{2} - E T\phi^{3} + {\lambda_{T}\over 4}\phi^{4}.
\label{effpot}
\end{equation}
The quantities $D$, $E$, $T_{o}$ and $\lambda_{T}$ must be determined
for each specific model. The presence of the cubic term ensures a
first-order phase transition.
A  minimum occurs, for a non-zero value of $\phi$, at the phase transition when all three terms in
the effective potential are roughly of the same order of magnitude, that is 
 $\phi \sim {E\over \lambda} T$.
Thus, roughly a first constraint of the value of the Higgs mass arises 
\begin{equation}
{E_{sph}\over T} \sim {m_{W}\over g^{2} T} \sim {\phi\over g T} \sim {g^{2}\over \lambda} \sim {m_{W}^{2}\over m_{H}^{2}},
\label{mWmH}
\end{equation}
where we have included only gauge bosons contributions to the effective potential so that $E\sim g^{3}$.
This implies that the electroweak phase transition will be sufficiently  of first order only for
small enough values of $m_{H}$ given the constraint of eq. (\ref{EsphT}).
In order to perform a more detailed calculation, we must include the
effect of all of the particles, and also higher-order corrections can be important.
However, the analysis is  constrained by the appearance of infrared divergences due to the massless gauge bosons in the symmetric phase. Resummation has been used 
to incorporate leading higher-order effects to one and two-loops in models
for electroweak baryogenesis \cite{dine, oespinosa}. The net effect of resummation in
the Standard Model is to weaken the strength of the phase transition. In addition, even with resummation the results of 1-loop calculations are unreliable for values of the Higgs field $\phi \lsi gT$.
In order to tackle this problem non-perturbative methods have been 
put forward.  An interesting
approach was introduced by Kajantie et al. \cite{farakos, kajantie, KLRS},
in which the main idea is to separate the perturbative and non-perturbative aspects
of the theory. The fact that there exists, at finite temperature, a
hierarchy of mass scales  can be used to construct effective 3D
theories for a given model. The perturbative component of the calculation, called dimensional reduction,
consists of constructing an effective 3D theory by matching 3D and 4D Green's functions of the
light degrees of freedom up to a certain order in the loop expansion.
At finite temperature the non-zero Matsubara modes have masses $\sim \pi T$, and  a further reduction can be performed generically, noting that some of the static modes in the theory, as a result of integrating out the non-static modes,
have acquired thermal masses proportional to a gauge coupling multiplied by the temperature $\sim g_{W(s)}T$. The
final effective theory contains a single light-scalar and transverse-gauge bosons. We emphasize that the whole procedure of dimensional reduction is
perturbative and free from infrared divergences. The  Lagrangian of the resulting purely bosonic theory is

\begin{equation}
L = {1\over 4} F_{ij}F_{ij} + {1\over 2} m_{3}^{2} \phi^{2} +
\lambda_{3}\phi^{4},
\label{L3D}
\end{equation}
 which facilitates lattice
simulations.
The lattice analysis, which
incorporates non-perturbative effects, translates the
constraint for a sufficiently strong first-order phase transition
 into an upper bound on the ratio of the 3D Higgs
coupling and the square 3D gauge coupling
\begin{equation}
x_{c} = {\lambda_{3}\over g_{3}^{2}} < 0.03 \cdots 0.04,
\label{xc}
\end{equation}
at the critical temperature.
The result of the analysis for the Standard Model, given the experimental constraints,  is that there
is no value of the Higgs mass that gives a sufficiently strong phase transition \cite{kajantie}.

\begin{figure}[tbh]

\vspace{-100pt}

\centerline{\hspace{-3.3mm}
\epsfxsize=8cm\epsfbox{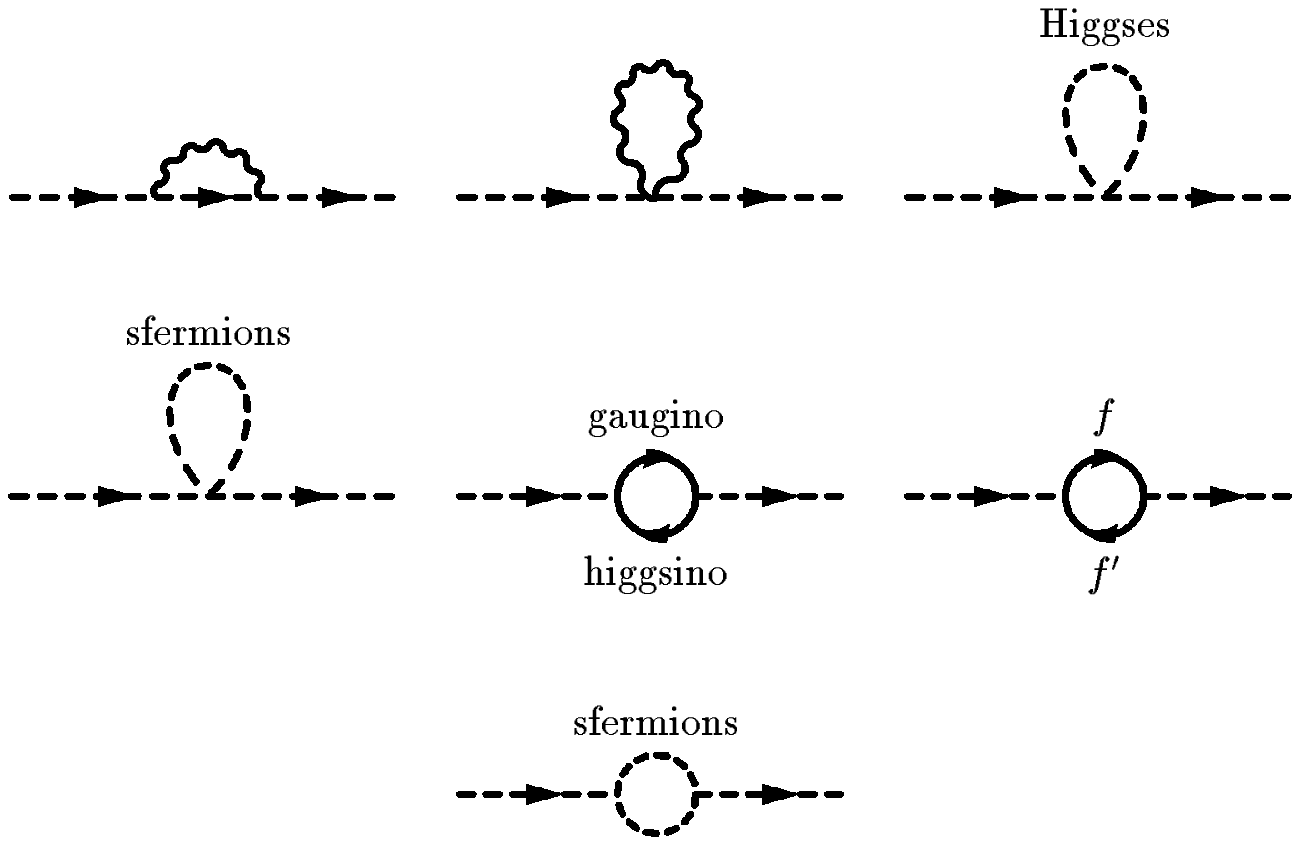}
\hspace{-1cm}
\epsfxsize=8cm\epsfbox{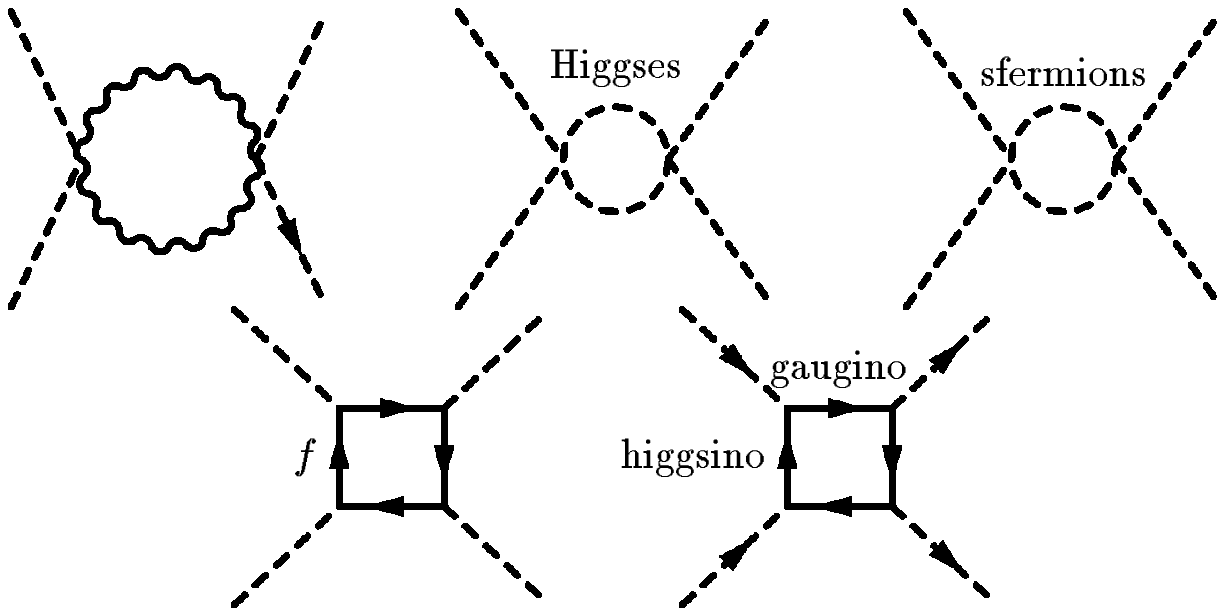}}

\vspace{-3.5cm}

\vspace*{-0.3cm}

\centerline{\hspace{1cm} (a) \hspace{6.5cm} (b)}

\caption[a]{{\small(a) Feynman diagrams contributing to the mass of the scalar Higgses and to
wave-function renormalization. (b) Diagrams contributing to the quartic Higgs couplings.}}
\end{figure}

\section{Dimensional Reduction in the MSSM}
In the Minimal Supersymmetric Standard Model (MSSM), it is also generic to have a single light scalar 
at the electroweak phase transition. Thus the effective theory is also
described by eq. (\ref{L3D}). Initially, perturbative 4D calculations
of the one-loop effective potential were used to analyse the strength of the phase
transition in this model \cite{zwirner1, zwirner2}.
The region of parameter space that was consistent with a
sufficiently strong phase transition was found to favour
low values of the ratio of the vacuum expectation values of 
the Higgs doublets $\tan\beta = {v_{2}\over v_{1}}$, and large 
values of the pseudoscalar mass $m_{A}$.

The procedure of dimensional reduction can be applied to
the MSSM and other extensions of the Standard Model \cite{Cline, Losada1, Laine}. For the MSSM the first stage of reduction will integrate out all
of the fermions (quarks, higgsinos, gauginos) and all non-zero Matusbara modes of the
bosonic fields (gauge bosons, sfermions, higgses). Figures 1a and 1b
show some of the diagrams that must be calculated for a one-loop
dimensional reduction for a two-point function and four-point function, respectively.

The general structure of the 3D couplings in terms of the
4D parameters is of the form
\begin{equation}
\lambda_{3} = T[\lambda - \beta_{\lambda}^{b} L_{b}- \beta_{\lambda}^{f} L_{f}
+ \rm{const.}].
\label{3Dlambda}
\end{equation}
where $\beta_{\lambda}^{b(f)}$ are the corresponding bosonic(fermionic)
$\beta$-functions. Similarly, for the mass terms we have the one-loop relation

\begin{equation}
\overline{m}_{3}^{2} = m^{2} + \eta_{m}(\mu)T^{2} - \beta_{m}^{b} L_{b}- \beta_{m}^{f} L_{f},
\label{3DmH}
\end{equation}
where $\eta_{m} \sim g_{W(s)}^{2}(\mu)$.
The 3D scalar and gauge couplings
are renormalization-group invariant. This implies that
a one-loop matching of the 3D coupling constants to
the physical parameters and the temperature suffices to
determine the strength of the phase transition, using
the constraint given by eq. (\ref{xc}). In order to complete the matching of the  3D parameters to 4D
physical parameters, the zero-temperature theory must be properly
renormalized.
We insist that the value of the critical temperature
does depend on a precise determination of the 3D  mass
parameter, which does get renormalized in the effective theory, see below;
 a two-loop calculation (in 4D) must be performed
even in the case of single light scalar at the phase transition.
However, as $x_{c}$ has only a weak dependence on the
temperature for values close to the critical temperature
of the phase transition for the allowed range of values of the
Higgs mass \cite{Losada2}, a two-loop calculation is not
necessary for the adequate suppression of the
sphaleron rate. The analysis of the dimensionally reduced theory at one loop
confirmed that, for the large-$m_{A}$ region of parameter space,
 a sufficiently strong first order phase transition occurs, with an upper bound for the Higgs mass of $m_{h} \lsi 80$ GeV. However, given
the current experimental limits on the Higgs mass, this region is excluded. 
Two 
possible ways of strengthening the phase transition, which can enlarge
the allowed region of parameter space, have been observed:

a) a fine-tuned scenario with a light right-handed stop \cite{Carena2}\footnote{Previous studies had shown that light stops strengthen the transition, but it is not possible to
make both of the stop fields light from the constraints
at zero temperature on the $\rho$ parameter \cite{zwirner2}.}. At  one-loop
the effect consists of enhancing the coefficient of the cubic term in eq. (\ref{effpot}), thus making the phase 
transition stronger:

\begin{equation}
E \simeq E_{SM} + {h_{t}^{3} \sin^{3}\beta\over2 \pi} \biggl(1 - {X_{t}^{2}\over m_{Q}^{2}}\biggr)^{3/2},
\label{EMSSM}
\end{equation}
where $h_{t}$, $X_{t}$, $m_{Q}$ are the top quarks Yukawa coupling, the stop
mixing parameter, and the soft SUSY-breaking mass of the third-generation
left squark doublet, respectively.
This scenario requires a soft SUSY-breaking mass for the right-handed stop $m_{U}^{2} < 0$, which tends to
cancels the positive temperature corrections to the thermal mass. However, this
may lead to physically unacceptable solutions at zero-temperature. In order to
avoid colour-breaking minima at zero temperature the following constraint
must be imposed \cite{Carena2}
\begin{equation}
-m_{U}^{2}\geq \biggl({m_{h}^{2} v^{2} g_{s}^{2}\over 12}\biggr)^{1/2},
\label{muC}
\end{equation}
which ensures that the physical minimum is deeper that the colour-breaking minimum.
This scenario could not be studied with the previous dimensional
reduction, as the perturbative expansion breaks down for values 
of the right-handed stop mass $m_{t_{R}} \lsi 190$ GeV \cite{Losada3}.

b) Higher-order QCD corrections from stops were also shown to be relevant by
affecting the value of the scalar field at the phase transition 
\cite{espinosa, deCarlos}.

We now direct our attention to
the analysis of the phase transition with a light right-handed stop.
The dimensional-reduction procedure has to be redone
 as the perturbative expansion starts to break down for smaller values of $m_{\tilde{t}_{R}}$. In addition new numerical simulations must also be performed; they will
have to take into account gluonic fields as they are not decoupled from the 
squarks in the 3D theory.
However, it is not so simple to obtain a constraint similar to that of eq.
(\ref{xc}) for the case of two light-scalar fields \cite{Laine3}.
A surprising result of the first perturbative analysis at two loops was the appearance
of a two-stage phase transition, where the Universe would reach the physical
vacuum having first gone through an intermediate colour-breaking
phase \cite{Laine2}.
The effective potential in the 3D theory, which reproduces the 4D results, can be studied in order to compare
with non-perturbative results as well\footnote{The first lattice analysis
seems to indicate that the results of the perturbative
calculations are conservative in the bounds that
are placed on the physical masses.}.

The final expression for the tree-level 3D potential is given by

\begin{eqnarray}
V_{3D} &=& \overline{m}_{H_{3}}^{2} H^{\dagger}H + \overline{\lambda}_{H_{3}}
(H^{\dagger}H)^{2} +  \overline{m}_{U_{3}}^{2} U^{\dagger}U \nonumber \\
&+& \overline{\lambda}_{U_{3}}
(U^{\dagger}U)^{2} + \overline{\gamma}_{3}(H^{\dagger}H)(U^{\dagger}U);
\label{V3D}
\end{eqnarray}
the  explicit expressions for the masses and couplings can be found
in ref. \cite{Losada3}.
A precise determination of the critical temperatures is necessary to ensure
the existence of a two-stage phase transition.
The most relevant quantities that define the critical temperatures
are the 3D mass parameters for the Higgs doublet and the right-handed
stop

\begin{equation}
\overline{m}_{H_{3}}^{2}(\mu) = \overline{m}_{H_{3}}^{2} + {1\over (16\pi^{2})} f_{2m_{H}}\log{\Lambda_{H_{3}}\over \mu},
\label{mH3mu2}
\end{equation}

\begin{equation}
\overline{m}_{U_{3}}^{2}(\mu) = \overline{m}_{U_{3}}^{2} + {1\over (16\pi^{2})} f_{2m_{U}}\log{\Lambda_{U_{3}}\over \mu},
\label{mU3mu2}
\end{equation}
where $\overline{m}_{H_{3}(U_{3})}^{2}$ have the form of eq. (\ref{3DmH}).
 
The expressions for the two-loop beta functions $f_{2m_{H}},f_{2m_{U}}$ for the mass parameters
have been given in ref. \cite{Laine2}. As mentioned there,
in order to fix the values of the parameters $\Lambda_{H_{3}}$ and $\Lambda_{U_{3}}$, we must employ  the two-loop  effective potential of the 4D theory
\footnote{In refs. \cite{Laine2, Laine3} an estimate of $\Lambda_{H_{3}} \sim
\Lambda_{U_{3}} \sim 7T$ was used.}. In addition two-loop effects can be numerically important because the scale of the couplings in the thermal contributions
to eq. (\ref{3DmH}) can only be fixed at the two-loop level.
The strategy we employ follows that of ref. \cite{KLRS}.
The idea is to use the 4D two-loop effective potential in order to fix
the scales in the 3D theory, and to use the 3D effective potential expressions
for the Higgs and stop fields
given in ref. \cite{Laine2} to analyse the phase transition 
We  calculate the unresummed two-loop effective potential
in order to include all 4D corrections to the mass parameters; resummation
is automatically  included in the calculation of the two-loop effective
potential in the 3D theory. We must also include the
contributions  to the two-loop effective potential of the static modes, which have been
integrated out  at the second stage  (includes the effects of resummation of the heavy fields)\footnote{ We derive the effective potential using the background fields $\phi$ and $\chi=\tilde{t}_{R\alpha} u^{\alpha}$, where we have chosen the unit vector in colour
space $u^{\alpha} = (1,0,0)$. The details of the calculation are found in ref. \cite{Losada3}.}.

\begin{figure}[ht]

\vspace{-2cm}

\centerline{\hspace{-3.3mm}
\epsfxsize=7cm\epsfbox{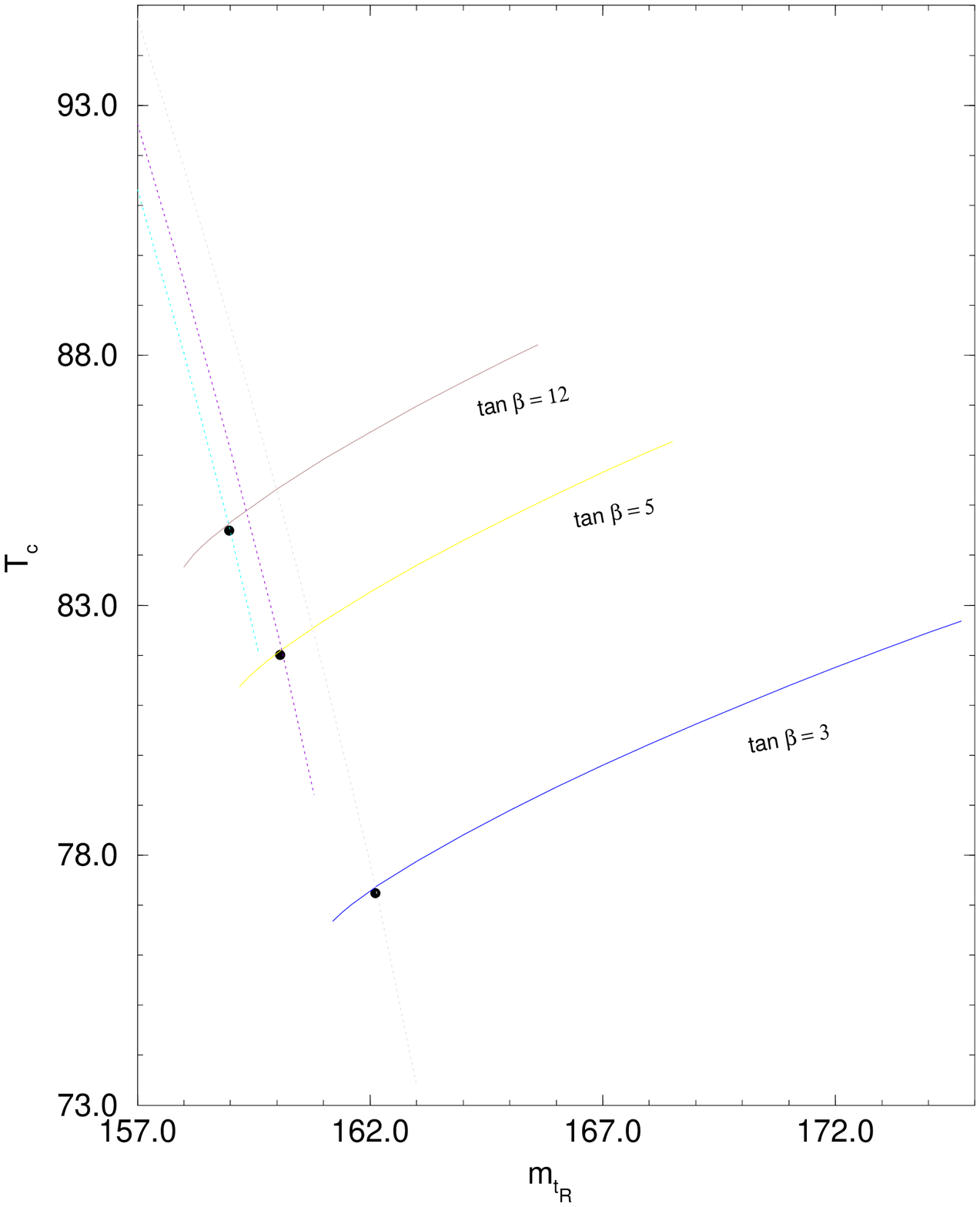}
\hspace{-.2cm}
\epsfxsize=7cm\epsfbox{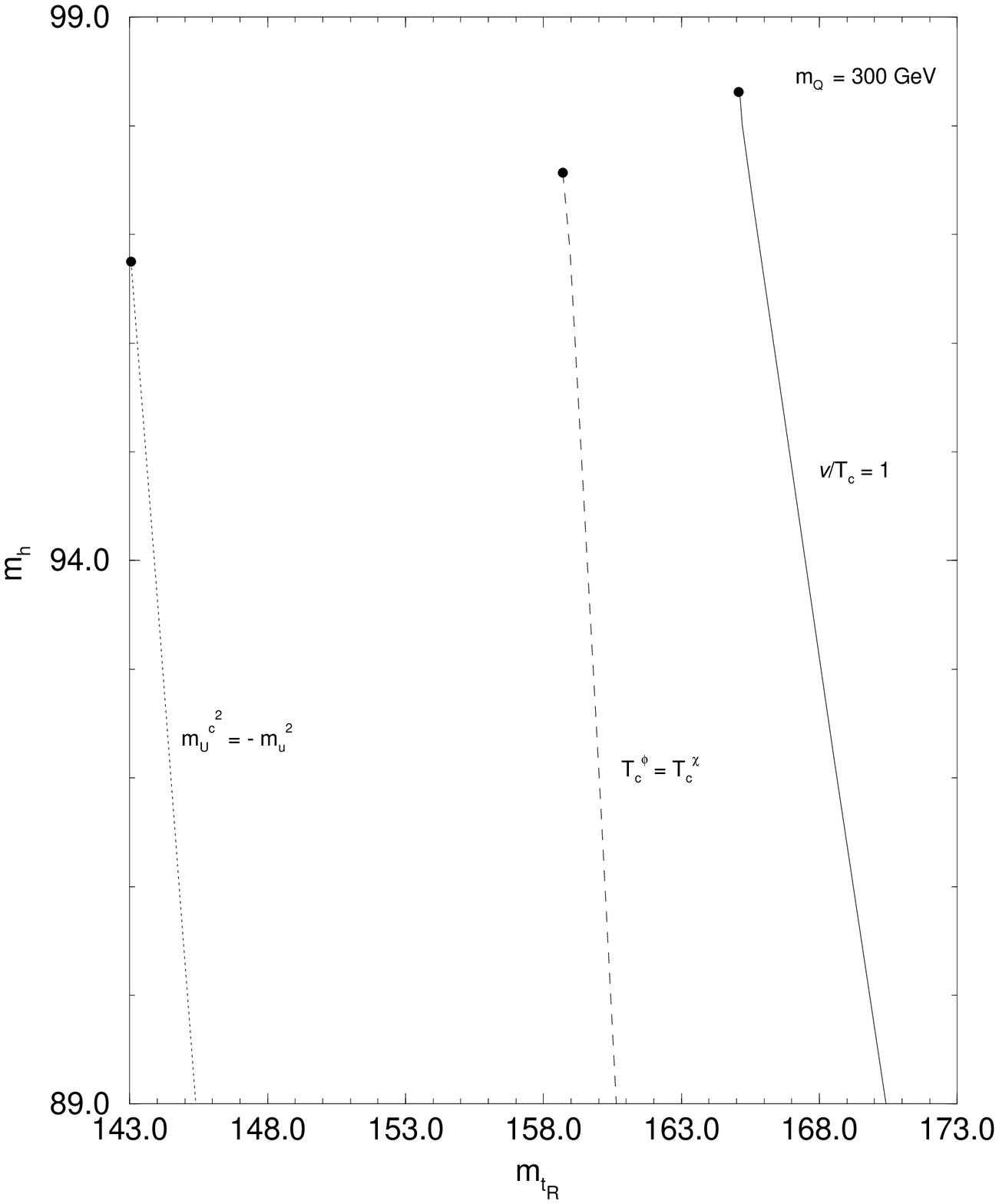}}

\vspace{-0.3cm}

\vspace*{-0.2cm}

\centerline{\hspace{1cm} (a) \hspace{6.5cm} (b)}

\caption[a]{{\small (a) Critical temperatures in the $\phi$ (solid) and $\chi$ (dotted) directions
as functions of $m_{\tilde{t}_{R}}$ for $\tan\beta =3,5,12$ and  $m_{Q} = 300$ GeV. (b)Allowed region in $m_{h}$--$m_{\tilde{t}_{R}}$ plane for $m_{Q}=300$ GeV. To the left of the solid line
there is a sufficiently strong first-order phase transition, to the right of the dotted line
the physical vacuum is absolutely stable. The dashed line separates the region for which
a two-stage phase transition can occur.} }
\end{figure}

Having done this we can now analyse the phase transition.
 In fig. 2a we
show the critical temperatures for the transitions in the $\phi$- and
$\chi$-directions as a function of the right-handed stop pole mass $m_{\tilde{t}_{R}}$, for $\tan\beta =3,5,12$. We find that, for  $m_{Q} \sim 300$ GeV, there
 still is a region in which a two-stage phase transition can occur. This region
is to the left of the crossing points of the curves.
With respect to the work of ref. \cite{Laine2} the structure of the
phase diagram is preserved, although it is slightly shifted towards
higher
values of the right-handed stop mass.
The total effect  does not substantially increase or decrease 
 the range of values of the right-handed stop mass for which a two-stage phase transition can occur. However, the exact location of this small range in the value
of $m_{\tilde{t}_{R}}$ depends on the value of the third-generation left-handed squark doublet mass. We note that the strength of the phase transition
has a weak dependence on the values of the scales that have been fixed in our calculation, and only slight differences
are observed with respect to previous analyses. 

The allowed region in parameter space is shown in fig. 2b, given the
current experimental limits on the Higgs mass \cite{moriond}. The region on the left of the solid line indicates
when a sufficiently strong first-order phase transition occurs. The
dotted line gives the condition for absolute stability of the
physical vacuum.  As explained above, to the left of this line the colour-breaking minimum is
lower than the physical one at zero-temperature. The dashed line is obtained
when the critical temperatures of the transitions in the $\phi$- and $\chi$-directions
are the same. A two-stage phase transition occurs to the left of the dashed line.

 \section{Conclusions}
We have performed a full two-loop dimensional reduction
of 4D MSSM parameters to the 3D couplings and masses of the effective theory.
In this way, we have fixed the scales appearing in the 3D mass terms
 that are due to the thermal polarizations and the
super-renormalizability of the 3D theory. The values of the
parameters $\Lambda_{H_{3}}$ and $\Lambda_{U_{3}}$ can vary 
significantly  for different values of the input parameters and
the particle content of the theory, thus modifying the critical temperatures of
the transitions. This effective theory can now be used for lattice simulations.
 We conclude that the allowed range of
masses is $m_{h} \lsi 110$ GeV and $m_{\tilde{t}_{R}} \lsi m_{t}$, in complete
agreement with previous results. We find that the phase structure diagram
still allows a possible two-stage phase transition for a small
range of values of $m_{\tilde{t}_{R}}$. This range of values is shifted
with respect to previous results. 
However,  whether or not the transition actually occurs must be explicitly checked.
Initial lattice analysis suggests that the second stage of the transition
is extremely strong and thus  this transition might  not
have taken place on cosmological time scales. Consequently,  this region of parameter space
for electroweak baryogenesis would be excluded.

\subsection*{Acknowledgments}
 I would like to thank the organizers of the Quarks98
conference for an interesting and enjoyable meeting.

}
\end{document}